\documentclass[conference]{IEEEtran}
\IEEEoverridecommandlockouts
\usepackage{cite}
\usepackage{amsmath,amssymb,amsfonts}
\usepackage{algorithm}
\usepackage{algorithmic}
\usepackage{graphicx}
\usepackage{textcomp}
\usepackage{xcolor}
\usepackage{color}
\usepackage{enumitem}
\usepackage{booktabs}
\usepackage{threeparttable}
\usepackage{multirow}
\usepackage{hyperref}
\usepackage{etoolbox}
\makeatletter
\patchcmd{\@makecaption}
  {\scshape}
  {}
  {}
  {}
\makeatletter
\patchcmd{\@makecaption}
  {\\}
  {.\ }
  {}
  {}
\makeatother
\def\tablename{Table}

\newtheorem{definition}{Definition}
  
\DeclareRobustCommand*{\IEEEauthorrefmark}[1]{%
    \raisebox{0pt}[0pt][0pt]{\textsuperscript{\footnotesize\ensuremath{#1}}}}
\begin{document}

% \title{HCL-DTI: Heterogeneous Contrastive Learning for Predicting Drug-Target Interatcion\\
\title{\huge A Heterogeneous Network-based Contrastive Learning Approach for Predicting Drug-Target Interaction\\
% {\footnotesize \textsuperscript{*}Note: Sub-titles are not captured in Xplore and
% should not be used}
% \thanks{Identify applicable funding agency here. If none, delete this.}
}

\author{
\IEEEauthorblockN{
Junwei Hu \IEEEauthorrefmark{1},
Michael Bewong \IEEEauthorrefmark{2,3},
Selasi Kwashie \IEEEauthorrefmark{3},
Wen Zhang \IEEEauthorrefmark{1}, Vincent M. Nofong \IEEEauthorrefmark{4},\\
Guangsheng Wu \IEEEauthorrefmark{5},
Zaiwen Feng \IEEEauthorrefmark{1,*}\thanks{*Correspondence: Zaiwen.Feng@mail.hzau.edu.cn}}
\IEEEauthorblockA{\IEEEauthorrefmark{1} College of Informatics, Huazhong Agricultural University, Wuhan, Hubei, China}
\IEEEauthorblockA{\IEEEauthorrefmark{2} School of Computing, Mathematics \& Engineering, Charles Sturt University, Wagga Wagga, NSW, Australia}
\IEEEauthorblockA{\IEEEauthorrefmark{3} Artificial Intelligence \& Cyber Futures Institute, Charles Sturt University, Bathurst, NSW, Australia}
\IEEEauthorblockA{\IEEEauthorrefmark{4} Faculty of Engineering, University of Mines and Technology, Tarkwa, Ghana}
\IEEEauthorblockA{\IEEEauthorrefmark{5} School of Mathematics and Computer Science, Xinyu University, China}
}
\maketitle

\begin{abstract}
Drug-target interaction (DTI) prediction is crucial for drug development and repositioning. Methods using heterogeneous graph neural networks (HGNNs) for DTI prediction have become a promising approach, with attention-based models often achieving excellent performance. However, these methods typically overlook edge features when dealing with heterogeneous biomedical networks. 
We propose a heterogeneous network-based contrastive learning method called \emph{HNCL-DTI}, which designs a heterogeneous graph attention network to predict potential/novel DTIs. 
Specifically, our \emph{HNCL-DTI} utilizes contrastive learning to collaboratively learn node representations from the perspective of both node-based and edge-based attention within the heterogeneous structure of biomedical networks. 
Experimental results show that \emph{HNCL-DTI} outperforms existing advanced baseline methods on benchmark datasets, demonstrating strong predictive ability and practical effectiveness.
The data and source code are available at \href{https://github.com/Zaiwen/HNCL-DTI}{https://github.com/Zaiwen/HNCL-DTI}.
\end{abstract}

\begin{IEEEkeywords}
drug-target interaction prediction, heterogeneous graph neural networks, heterogeneous biomedical networks, heterogeneous graph attention network, contrastive learning
\end{IEEEkeywords}

\section{Introduction}\label{Intro}
% Predicting drug-target interaction (DTI) is a crucial aspect of biomedical research. It helps scientists understand how drugs work in the body and design new therapies. 

% For example, (1) \textbf{\emph{Drug Discovery:}} By identifying potential targets, researchers can prioritize which molecules to investigate further as medications. This can significantly accelerate the drug discovery process, saving time and resources. (2) \textbf{\emph{Drug Repurposing:}} DTI prediction can help identify existing drugs that might be effective for new uses. This is a more efficient approach than developing entirely new drugs from scratch.
% (3) \textbf{\emph{Precision Medicine:}} Understanding how drugs interact with specific targets allows for the development of personalized treatments tailored to an individual's genetic makeup. This can lead to more effective therapies with fewer side effects.

% Overall, accurate DTI prediction has the potential to revolutionize drug development and treatment, leading to faster creation of more effective and personalized medicines~\cite{yue2020graph}.

% Drug target interaction (DTI) refers to the specific binding between a drug molecule and a biomolecule within the body. This biomolecule, often called the drug target, is usually a protein, enzyme, receptor, or other molecule that plays a crucial role in a biological process. When a drug interacts with its target, it can alter that process in some way, leading to the drug's therapeutic effect.

Predicting drug-target interactions (DTIs) is significant for accelerating drug repositioning, given that approximately 75\% of drugs can be repurposed ~\cite{nosengo2016can}. 
Identifying DTIs using traditional experimental methods is time-consuming and expensive.  
Recently, heterogeneous graph neural network (HGNN)-based DTI prediction methods have attracted widespread attention from both academia and industry.
However, most existing HGNN-based methods rely solely on chemical and genomic data to predict DTIs~\cite{gao2018interpretable, huang2021moltrans, nguyen2021graphdta, chu2021dti}, ignoring useful pharmacological and phenotypic information \emph{i.e.}, diseases and side-effects~\cite{adasme2021structure}.
The intricate connections among drugs, targets, diseases, and side-effects offer valuable insights for understanding DTIs at a system level. Therefore, incorporating diverse biological data may aid in predicting DTIs and further facilitate drug repurposing.
% The interrelationships among biological entities such as drugs, targets, diseases, and side-effects contain rich semantic information and provide a system-level understanding of DTIs. Therefore, integrating heterogeneous biological data may aid in predicting DTIs and further facilitate drug repurposing.

HGNN-based techniques have attracted increasing attention due to their successful and significant performance on prediction tasks in the biomedical domain. 
% NRLMF-$\beta$~\cite{ban2019nrlmfbeta} models the probability of drug-target interactions through logical matrix factorization. 
DTINet~\cite{luo2017network} accurately encodes the topological information of each node in a heterogeneous network into its learned node representations. 
NeoDTI~\cite{wan2019neodti} incorporates a network topology preservation mechanism to align learned drug and target embeddings with the underlying network structure.
GCN-DTI~\cite{zhao2021identifying} first leverages a graph convolutional network to learn the feature for each drug-target pair and then uses a deep neural network to identify DTIs. 
% In recent years, graph learning methods have achieved excellent performance on tasks such as node classification and link prediction. 
SGCL-DTI~\cite{li2022supervised} utilizes drug-traget pairs (DTPs) to perform graph contrastive learning, improving the consistency of node representations linked to labeled data both semantically and topologically.
HGAN~\cite{li2022heterogeneous} applies a graph attention network to capture complex structures and rich semantics in the biological heterogeneous graph for DTI prediction.
Although these methods show significant advantages, relational information is often ignored in their approaches when dealing with heterogeneous biomedical networks. However, explicitly incorporating relational information into the model can be used to further improve the performance of prediction tasks.   
%which also leaves a potential space to incorporate relational information into the model to improve the performance of prediction tasks.

\begin{figure*}[htbp]
    \centering  
    \includegraphics[width=1\linewidth]{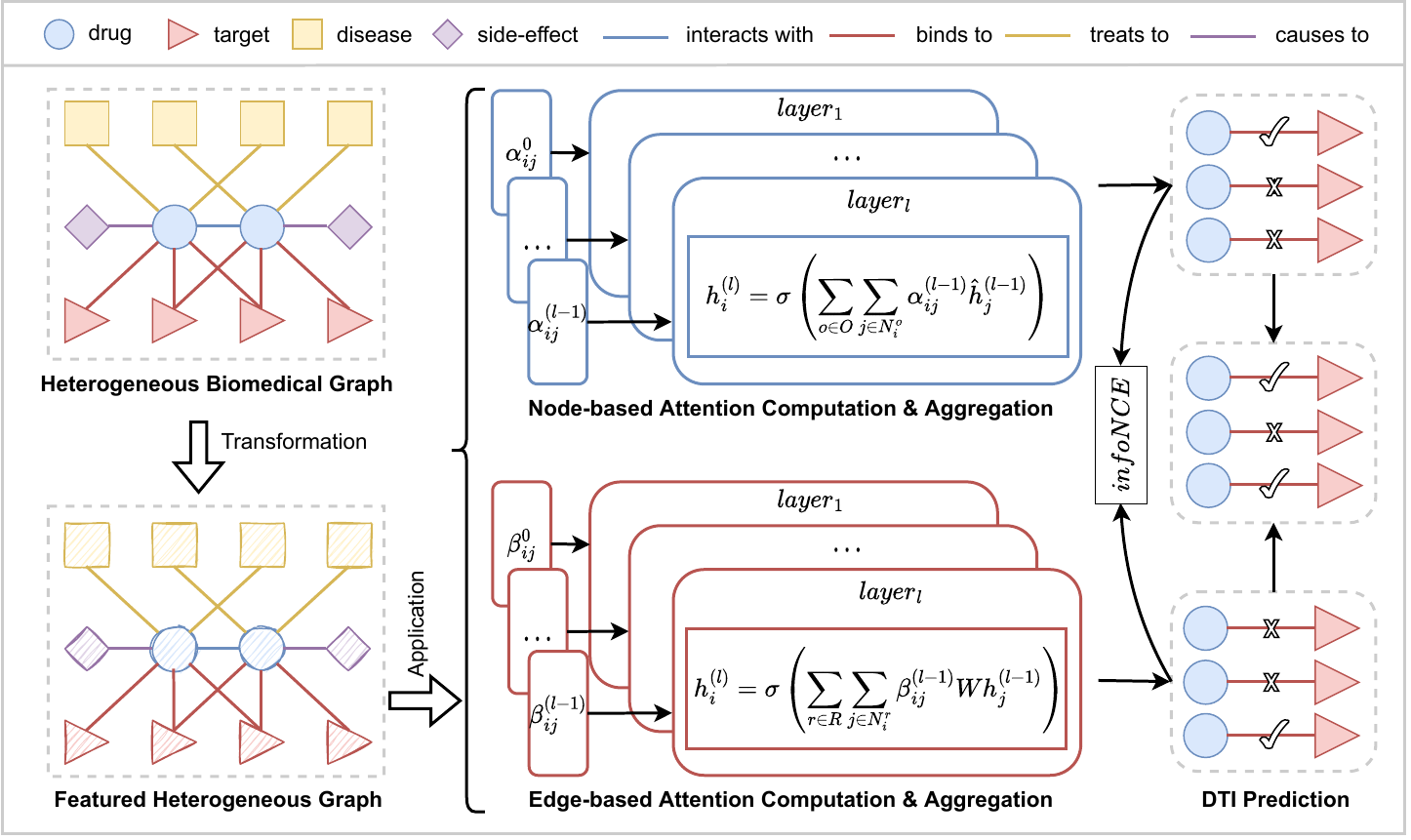}
    \vspace{-4ex}
    \caption{Our proposed DTI prediction framework.}
    \label{HNCL-DTI}
    \vspace{-2ex}
\end{figure*}

We propose a DTI prediction method called \emph{HNCL-DTI} as shown in Fig.~\ref{HNCL-DTI}. Specifically, we utilize two different attention mechanisms to update the embedding representation of nodes. One is the node-based attention mechanism, which calculates the attention coefficient by performing dot-product on different types of node features, and the other is the edge-based attention mechanism, which introduces features of different relationships to learn the attention coefficient through a feed-forward neural network. Then, we use positive and negative sample pairs to train the objective function in contrastive learning to improve the representation consistency of the embedding vector. In summary, the contributions of this paper can be summarized as follows:

\begin{itemize}
\item A heterogeneous network-based contrastive learning approach for DTI prediction is proposed, which provides a novel idea for learning network structure information in heterogeneous graph neural network models.

\item We adopt two different attention mechanisms (\emph{i.e.}, node-based and edge-based) to update the feature vectors of nodes separately, and then apply contrastive learning to collaboratively learn node representations from both node-based and edge-based attention perspectives.

\item We evaluate the performance of \emph{HNCL-DTI} on two heterogeneous biomedical datasets. Experimental results show that \emph{HNCL-DTI} achieves better performance than 9 advanced baseline methods. And our case study demonstrates the effectiveness of \emph{HNCL-DTI} in real scenarios.
\end{itemize}

The rest of this paper is organized as follows. Section~\ref{RW} reviews the related work. Section~\ref{Pre} and ~\ref{Method} introduces the basic definitions and overall design of our model respectively. Then, Section~\ref{Exper} discusses the experimental setup and results in detail. Finally, we conclude this paper in Section~\ref{Con}.

\section{Related work}\label{RW}
Drug-target interaction (DTI) refers to the binding of a drug molecule to a target, typically a protein, where the drug interacts with the target to treat diseases. Enhancing the accuracy of DTI predictions could accelerate drug development.

Traditional computer-based methods for DTI prediction have limitations and are time-consuming. Docking-based approaches~\cite{morris2009autodock4}, which rely on the 3D structure of the target protein, and ligand-based methods, which utilize known drug interaction patterns, are examples of these early techniques~\cite{keiser2007relating}. Both methods have constraints and can be computationally expensive.

Advances in biological and chemical technologies have led to a surge in available data, including drug-drug and drug-target interactions, which can improve the accuracy of DTI predictions. To leverage this expanding dataset, network-based approaches have been developed that employ graph-based techniques to characterize the properties of drugs and targets to improve DTI predictions.
MultiDTI~\cite{zhou2021multidti} leverages a heterogeneous network-based framework to integrate similarity and network-based approaches for joint representation learning.
IMCHGAN ~\cite{li2021imchgan} employs a two-level graph attention network to extract latent feature representations for drugs and targets from multiple interconnected networks. These learned representations are subsequently used in an inductive matrix completion model to predict DTIs.

\section{Preliminaries}\label{Pre}
This section presents formal definitions of the terms used in this paper. 

\begin{definition}[Heterogeneous Network]
A heterogeneous network can be formulated as a network $G = (V, E, \phi, \psi)$, where $V$, $E$ denote the node type and edge type set respectively, and $\phi$, $\psi$ represent the node type mapping function $\phi : V \rightarrow O$ and edge type mapping function $\psi : E \rightarrow R$ respectively. If $|R| + |O| > 2$, the network is \textbf{ \emph{heterogeneous}}. $\square$
\end{definition}

We show a toy example of a heterogeneous network in Fig.~\ref{HNCL-DTI}, which contains four relationships (\emph{i.e.}, \emph{interacts with}, \emph{binds to}, \emph{treats to}, \emph{causes to}) and four node types (\emph{i.e.}, \emph{drug}, \emph{target}, \emph{disease}, \emph{side-effect}).

\begin{definition}[Heterogeneous Graph Neural Network]
Given a heterogeneous network $G$, the purpose of a HGNN is to derive a representation vector $\textbf{\emph{h}}_v^{(l)} \in \mathbb{R}^{d_l}$ for each node $v$ via $l$-layer transformations, utilizing both the graph structure and the initial node feature $\textbf{\emph{h}}_v^{(0)} \in \mathbb{R}^{d_0}$, where $d_0$ and $d_l$ represent the featured dimensionality of node $v$ in layer 0 and $l$ respectively. These learned representations can be applied to different application tasks such as drug-drug interaction (DDI), protein-protein interaction (PPI), and drug-disease association (DDA) prediction. $\square$
\end{definition}

\iffalse
For example, 
%when we get the final embedding representation of nodes, 
consider a final embedding representation of nodes
%there is a drug node $i$ with a node representation 
containing a \emph{drug} node $i$ with the node representation
$\textbf{\emph{h}}_{i}^{(l)}$ and a \emph{target} node $j$ with a node representation $\textbf{\emph{h}}_{j}^{(l)}$. We determine whether they are connected through the following formula:
\begin{equation}
Pred(i, j) = sigmoid(\textbf{\emph{h}}_{i}^{(l){\mathsf{T}}}\cdot \textbf{\emph{h}}_{j}^{(l)})
\end{equation}
If $\mathsf{True}$ is returned, it is considered to be a positive drug-target pair (DTP), otherwise it is considered to be a negative pair.
\fi

\section{Method}\label{Method}
\subsection{Node-based Attention Mechanism}\label{NbAM}
For each heterogeneous biomedical network, different types of nodes may have different feature distributions. First, we assign a unique linear projection to each node type to maximize the simulation of distribution differences. Then, based on the projected features, we use the scaled dot-product to calculate an attention weight between two nodes connected by an edge. Finally, the feature update of the node will aggregate different types of weighted neighbor information. The above process is mainly divided into the following three steps:

\subsubsection{Feature Transformation} For each node $i \in V$, there is a specific node type $\phi(i) \in O$ with the feature $\textbf{\emph{h}}_i^{(l)}$, and $\textbf{\emph{h}}_i^{(l)}$ will be transformed through the formula:
\begin{equation}
    \hat{\textbf{\emph{h}}}_i^{(l)} = {\rm Linear}_{\phi(i)} (\textbf{\emph{h}}_i^{(l)})
\end{equation}
where ${\rm Linear}_{\phi(i)}$ is a projection function of node type $\phi(i)$.

\subsubsection{Attention Computation} Based on transformed node features, a node-based attention coefficient is computed for each link edge $e = \langle i, j \rangle \in E$, where $i, j \in V$:
\begin{equation}
    \alpha_{ij}^{(l)} = \frac{\exp(a(\hat{\textbf{\emph{h}}}_{i}^{(l)\mathsf{T}} \cdot \hat{\textbf{\emph{h}}}_{j}^{(l)}))}{\sum\limits_{ o \in O} \sum\limits_{k \in N_i^o}\exp(a(\hat{\textbf{\emph{h}}}_{i}^{(l)\mathsf{T}} \cdot \hat{\textbf{\emph{h}}}_{k}^{(l)}))} 
\end{equation}
where $a$ is the scaling factor, which serves as an adaptive scale for attention, $N_i^o$ denote the neighborhood of node $i$ within node type $o$. 

\subsubsection{Neighborhood Aggregation} Then the neighborhood information aggregation for node $i$ can be performed as:
\begin{equation}
    \textbf{\emph{h}}_i^{(l+1)} = \sigma \left ( \sum\limits_{o \in O} \sum\limits_{j \in N_i^o} \alpha_{ij}^{(l)} \hat{\textbf{\emph{h}}}_j^{(l)} \right )
\end{equation}
where $\sigma$ is an activation function ELU~\cite{clevert2015fast}.
To improve model expressiveness, we adopt a multi-head attention mechanism similar to GAT. In particular, we execute $T$ independent attention operations as described in formula (3) and average the outputs to produce the final representation. The corresponding update rule is:
\begin{equation}
    \textbf{\emph{h}}_i^{(l+1)} = \frac{1}{T} \sum\limits_{t=1}^T \textbf{\emph{h}}_{i,t}^{(l+1)}
\end{equation}

\subsection{Edge-based Attention Mechanism}\label{EbAM}
To represent heterogeneous biomedical networks, we adopt the attention mechanism of a feed-forward neural network to incorporate edge type information into the attention computation. To be specific, we allocate a $d$-dimensional embedding $\textbf{\emph{h}}_r^{(l)}$ for each edge type $r=\psi(e) \in R$ at each layer. The edge-based attention coefficient is computed for each edge  $e = \langle i, j \rangle \in E$ by the formula:
\begin{equation}
    \beta_{ij}^{(l)} = \frac{\exp(\tau(\Vec{\textbf{a}}^\mathsf{T}(\textbf{\emph{Wh}}_{i}^{(l)}||\textbf{\emph{Wh}}_{j}^{(l)}||\textbf{\emph{W}}_r\textbf{\emph{h}}_{\psi(\langle i, j \rangle)}^{(l)})))}{\sum\limits_{r \in R} \sum\limits_{k \in N_i^r}\exp(\tau(\Vec{\textbf{a}}^\mathsf{T}(\textbf{\emph{Wh}}_{i}^{(l)}||\textbf{\emph{Wh}}_{k}^{(l)}||\textbf{\emph{W}}_r\textbf{\emph{h}}_{\psi(\langle i, k \rangle)}^{(l)})))} 
\end{equation}
where $\tau$ denotes an activation function LeakyReLU~\cite{maas2013rectifier}, $\Vec{\textbf{a}}$ is the weight vector, $||$ is a concatenation operation, $\textbf{\emph{W}}$ and $\textbf{\emph{W}}_r$ represent a weight matrix and a learnable matrix to transform edge type embeddings respectively.

Then, the updated representation of target node $i$ can be obtained through the aggregation of neighborhood information as follows:
\begin{equation}
    \textbf{\emph{h}}_i^{(l+1)} = \sigma \left ( \sum\limits_{r \in R} \sum\limits_{j \in N_i^r} \beta_{ij}^{(l)} \textbf{\emph{Wh}}_j^{(l)} \right )
\end{equation}
similarly, the multi-head attention mechanism can be employed following the formula (4).

\subsection{Contrastive Learning}\label{CL}
Contrastive learning has proven to be highly effective in various graph learning tasks. Inspired by that, we propose using contrastive learning between node-based and edge-based attention aggregations to enhance the model’s representation learning capability. This approach maximizes the agreement of node representations learned across different views while capturing different heterogeneous information.

For this module, we consider the same nodes as positive samples and different nodes as negative samples in both node-based and edge-based attention aggregations. The node representations obtained through formula (3) for node-based attention aggregation and formula (6) for edge-based attention aggregation are denoted as $\textbf{\emph{H}}^{node}$ and $\textbf{\emph{H}}^{edge}$, respectively. Then, we have the following cross-view contrastive loss with InfoNCE~\cite{oord2018representation} as:
\begin{equation}
    loss_{CL} = -\sum\limits_{i \in V}\log\frac{\exp(c(\textbf{\emph{H}}^{node}_i, \textbf{\emph{H}}^{edge}_i)/\omega)}{\sum\limits_{j \in V}\exp(c(\textbf{\emph{H}}^{node}_i, \textbf{\emph{H}}^{edge}_j)/\omega)}
\end{equation}
where $c$ denotes the cosine similarity function, and $\omega$ represents the tunable temperature hyperparameter to adjust the scale for softmax.  In this way, the node-based and edge-based attention views are allowed to collaboratively supervise each other, thus enhancing the node representation learning.

We ultimately use the outputs of both the node-based and edge-based attention aggregation to
obtain the final and unified node representations $\textbf{\emph{H}}$ %\in \mathbb{R}^{n \times d}$
through an average pooling operation:
\begin{equation}
    \textbf{\emph{H}} = \frac{1}{2}(\textbf{\emph{H}}^{node}+\textbf{\emph{H}}^{edge})
\end{equation}
% These combined representations are then used for the DTI prediction downstream application.

\subsection{Model Learning}\label{ML}
We present the objective function to train our model to learn the final node representations. Following the DTI prediction task, we can train our model using binary cross-entropy loss function through negative sampling:
\begin{equation}
    loss_{DTI} = -\sum_{\substack{(i,j)\in \Omega,\\ (i',j')\in \Omega^-}}\log \delta(c(\textbf{\emph{H}}_i^\mathsf{T}, \textbf{\emph{H}}_j)\cdot (-c(\textbf{\emph{H}}_{i'}^\mathsf{T}, \textbf{\emph{H}}_{j'})))
\end{equation}
where $\delta$ is the sigmoid function, $c$ can be any vector similarity measure function, $\Omega$ represents the set of positive DTPs and  $\Omega^-$ denotes the set of negative DTPs randomly selected from unconnected node pairs.

Finally, we combine the DTI prediction loss with the contrastive learning loss to jointly optimize our model:
\begin{equation}
    loss = loss_{DTI} + \gamma \cdot loss_{CL}
\end{equation}
where $\gamma$ is the hyperparameter for tuning the importance of contrastive learning.

\section{Experiments}\label{Exper}
\subsection{Experimental Settings}
\subsubsection{Datasets}

There are two heterogeneous biomedical networks (HBNs) used to test our model \emph{HNCL-DTI}. $\textbf{HBN-A}$, a benchmark dataset introduced in DTINet~\cite{luo2017network}, is widely used for the DTI prediction task involving heterogeneous biological data.
The dataset
% , gathered from public resources~\cite{davis2013comparative, keshava2009human, knox2010drugbank, kuhn2010side}, 
includes 12015 bioentities categorized as 708 drugs (D), 1512 targets (T), 5603 diseases (I), and 4192 side-effects (S), and six types of connections (D-T, D-D, D-I, D-S, T-T, T-I) with a total of 1895445 connections. $\textbf{HBN-B}$ is an extension of HBN-A and was created by HGAN~\cite{li2022heterogeneous}, 
% It is gathered from a list of resources~\cite{wishart2018drugbank, kuhn2016sider, davis2021comparative, szklarczyk2021string}, 
consisting of 15322 bioentities and 5126875 interactions between them. The statistical information and resources about HBN-A and HBN-B are summarized in Table~\ref{HBNs}.

\begin{table}[ht]
\caption{The statistics and resources of HBN-A and HBN-B datasets}
\label{HBNs}
\resizebox{1\linewidth}{!}{
\begin{tabular}{@{}ccccc@{}}
\toprule
Network & Node Type & Number & Edge Type & Number \\ \midrule
\multirow{6}{*}{HBN-A} &  &  & drug-target & 1923 \\
 & drug & 708 & drug-drug & 10036 \\
 & target & 1152 & drug-disease & 199214 \\
 & disease & 5603 & drug-side-effect & 80164 \\
 & side-effect & 4192 & target-target & 7363 \\
 &  &  & target-disease & 1596745 \\ \midrule
\multirow{6}{*}{HBN-B} &  &  & drug-target & 8750 \\
 & drug & 2214 & drug-drug & 1091870 \\
 & target & 1968 & drug-disease & 542970 \\
 & disease & 7205 & drug-side-effect & 104629 \\
 & side-effect & 3935 & target-target & 456592 \\
 &  &  & target-disease & 2922064 \\ \bottomrule
\end{tabular}
}
\vspace{-2ex}
\end{table}

% \begin{table}[htbp]
% \resizebox{1\linewidth}{!}{
% \begin{tabular}{@{}cccccc@{}}
% \toprule
% \multirow{2}{*}{Types} & \multirow{2}{*}{Items} & \multicolumn{2}{c}{Hetero-A} & \multicolumn{2}{c}{Hetero-B} \\ \cmidrule(l){3-6} 
%  &  & Numbers & Resources & Numbers & Resources \\ \midrule
% \multirow{4}{*}{Node} & Drug (D) & 708 & DrugBank & 1519 & DrugBank \\
%  & Target (T) & 1152 & HPRD & 1025 & DrugBank,TTD,PharmGKB \\
%  & Disease (I) & 5603 & CTD & 1229 & ClinicalTrials.gov \\
%  & Side-effect (S) & 4192 & SIDER & 12904 & MetaADEBD,CTD,SIDER,OFFSIDES \\
% \multirow{6}{*}{Edge} & D-T & 1923 & DrugBank & 6744 & DrugBank,TTD,PharmGKB \\
%  & D-D & 10036 & DrugBank & 290836 & DrugBank \\
%  & D-I & 199214 & CTD & 6677 & ClinicalTrials.gov \\
%  & D-S & 80164 & SIDER & 382041 & MetaADEBD,CTD,SIDER,OFFSIDES \\
%  & T-T & 7363 & HPRD &  &  \\
%  & T-I & 1596745 & CTD &  &  \\ \bottomrule
% \end{tabular}
% }
% \end{table}

\subsubsection{Evaluation metrics} There are six metrics used to evaluate the model, including Recall, Precision, F1 score, the area under the receiver operating characteristic curve (AUC), the area under the precision-recall curve (AUPR), and the matthews correlation coefficient (MCC). In particular, the value range of MCC is between -1 and 1, a metric used to evaluate the performance of binary classification models. The value range of other metrics is between 0 and 1.

\subsubsection{Implementation details} A 10-fold cross-validation strategy is used to test known DTPs as well as randomly sampled unknown DTPs. For each fold, 90\%  is randomly sampled as the training set with equal number of positive and negative samples, and the remaining samples are employed as the test set. For \emph{HNCL-DTI}, the Adam is utilized as the optimizer with weight decay rate $1\times10^{-4}$ and learning rate $5\times10^{-3}$. The early stopping with the patience of 40 is utilized and the epoch is set to 300. By default, the hyperparameter $\gamma$, number of layers, dimensionality of embeddings, and number of attention heads are set to 0.01, 2, 128, and 8, respectively.
% For the settings of other parameters, see Section~\ref{Para}.
% The dimensions of the projected and hidden features are set to 128 and 32, respectively.

\subsubsection{Baselines} There are 9 advanced models (\emph{i.e.}, 6 biomedical methods and 3 heterogeneous graph embedding methods) with their default parameter settings as baseline methods:
\begin{itemize}
    \item \textbf{DTINet}~\cite{luo2017network} leverages a compact feature learning algorithm to generate low-dimensional vector representations that encode the topological features of each node in the network.
    \item \textbf{NeoDTI}~\cite{wan2019neodti} ensures that the learned representations of drugs and targets accurately reflect the original network structure by preserving network topology.
    \item \textbf{MultiDTI}~\cite{zhou2021multidti} integrates similarity-based and network-based approaches within a unified heterogeneous network framework to create joint representations.
    \item \textbf{IMCHGAN}~\cite{li2021imchgan} incorporates a two-level neural attention mechanism approach to derive latent feature representations for drugs and targets from the DTI heterogeneous graph.
    \item \textbf{SGCL-DTI}~\cite{li2022supervised} adopts a co-contrastive learning strategy for DTI prediction by contrasting the topology structures and semantic features of the DTP network.
    \item \textbf{HGAN}~\cite{li2022heterogeneous} utilizes a heterogeneous graph attention network to capture the complex structures and rich semantics for DTI prediction.
    \item \textbf{RGCN}~\cite{schlichtkrull2018modeling} is an extension of GCN designed for relational graphs (containing multiple edge types).
    \item \textbf{HGT}~\cite{hu2020heterogeneous} designs a novel strategy to calculate the mutual attention between source and target nodes.
    \item \textbf{Simple-HGN}~\cite{lv2021we} introduces a heterogeneous graph attention mechanism to compute attention scores between nodes.
\end{itemize}

\subsection{Effectiveness}
To evaluate the performance of models, we conduct a 10-fold cross-validation on the datasets and show the average of 10-fold results. We compare \emph{HNCL-DTI} with 9 baseline methods on the DTI prediction task. The performance of these models on two datasets (\emph{i.e.}, HBN-A and HBN-B) is shown in Table~\ref{HBN-A} and ~\ref{HBN-B}. 

It is obvious that \emph{HNCL-DTI} is significantly better than other baselines in all six evaluation metrics on HBN-A dataset, with improvements of 0.0175, 0.0217, 0.0173, 0.0358, 0.0621, and 0.1267 for AUC, AUPR, Precision, Recall, F1 score, and MCC, respectively. On the dataset HBN-B, it is noticed that \emph{HNCL-DTI} has the highest performance in four of six evaluation metrics over all baselines and the second highest performance in remaining two evaluation metrics. The results indicate the superiority of \emph{HNCL-DTI} in DTI prediction.

Compared with biomedical methods, \emph{HNCL-DTI} can capture relationship information in the heterogeneous biomedical graph and predict DTIs in an end-to-end manner. However, biomedical methods focus more on node (\emph{i.e.}, drug and target) features, while ignoring the richness of edge (\emph{i.e.}, interactions between them) information, which may lead to limited expressiveness of the model in capturing complex relationships. 
In addition, some of them separate the process of learning drug and target features from the prediction task, which can result in learned representations that are not optimally suited for DTI prediction.

Compared with heterogeneous embedding methods, \emph{HNCL-DTI} can learn representations of drug and target more deeply. However, heterogeneous embedding methods can also serve as an effective auxiliary tool for DTI prediction.

\begin{table}[ht]
\caption{The DTI prediction comparison results of HNCL-DTI with 9 baseline methods on dataset HBN-A. }
\label{HBN-A}
\begin{threeparttable} 
\resizebox{1\linewidth}{!}{
\begin{tabular}{@{}lcccccc@{}}
\toprule
Methods & AUC & AUPR & Precision & Recall & F1 & MCC \\ \midrule
DTINet & 0.8838 & 0.9024 & 0.8677 & 0.7369 & 0.7967 & 0.6316 \\
MultiDTI & 0.8951 & 0.9126 & 0.8253 & 0.7882 & 0.7898 & 0.6007 \\
NeoDTI & 0.8690 & 0.8821 & 0.7959 & 0.7904 & 0.7918 & 0.5825 \\
IMCHGAN & 0.8862 & 0.9000 & \underline{0.8747} & 0.7349 & 0.7982 & 0.6384 \\
SGCL-DTI & 0.7462 & 0.7910 & 0.7415 & 0.6280 & 0.6768 & 0.4193 \\
HGAN & \underline{0.9391} & 0.9303 & 0.7922 & \underline{0.8693} & 0.7944 & 0.5807 \\ \midrule
RGCN &  0.9032 & 0.9024 & 0.8606 & 0.7664 & 0.8108 & 0.6462 \\
HGT & 0.8486 & 0.8126 & 0.7883 & 0.8059 & 0.7970 & 0.5897 \\
Simple-HGN & 0.9333 & \underline{0.9392} & 0.8260 & 0.8471 & \underline{0.8364} & \underline{0.6690} \\ \midrule
HNCL-DTI & \textbf{0.9566} & \textbf{0.9609} & \textbf{0.8920} & \textbf{0.9051} & \textbf{0.8985} & \textbf{0.7957} \\ \bottomrule
\end{tabular}
}
\begin{tablenotes}    
     \item \emph{Note: The highest and second highest results are in bold and underlined, respectively.}
\end{tablenotes} 
\end{threeparttable}
% \vspace{-2ex}
\end{table}

\begin{table}[ht]
\caption{The DTI prediction comparison results of HNCL-DTI with 9 baseline methods on dataset HBN-B.}
\label{HBN-B}
\begin{threeparttable} 
\resizebox{1\linewidth}{!}{
\begin{tabular}{@{}lcccccc@{}}
\toprule
Methods & AUC & AUPR & Precision & Recall & F1 & MCC \\ \midrule
DTINet & 0.8239 & 0.8378 & 0.7900 & 0.6727 & 0.7265 & 0.4996 \\
MultiDTI & 0.8783 & 0.8883 & 0.8447 & 0.7647 & 0.8004 & 0.6254 \\
NeoDTI & 0.9072 & \underline{0.9078} & 0.8176 & 0.8387 & 0.8272 & 0.6488 \\
IMCHGAN & 0.8706 & 0.8863 & 0.8510 & 0.7427 & 0.7930 & 0.6179 \\
SGCL-DTI & 0.6945 & 0.7649 & 0.7706 & 0.5269 & 0.6255 & 0.3895 \\
HGAN & \underline{0.9247} & 0.9071 & 0.8165 & \textbf{0.8763} & \underline{0.8326} & \underline{0.6845} \\ \midrule
RGCN & 0.7624 & 0.8048 & \textbf{0.8644} & 0.5754 & 0.6909 & 0.5148 \\
HGT & 0.8154 & 0.8203 & 0.7442 & 0.7198 & 0.7318 &  0.4726 \\
Simple-HGN & 0.8841 &  0.8851 & 0.8145 & 0.7469 & 0.7793 & 0.5789 \\ \midrule
HNCL-DTI & \textbf{0.9290} & \textbf{0.9275} & \underline{0.8627} & \underline{0.8454} & \textbf{0.8540} & \textbf{0.7111} \\ \bottomrule
\end{tabular}
}
\begin{tablenotes}    
     \item \emph{Note: The highest and second highest results are in bold and underlined, respectively.}
\end{tablenotes} 
\end{threeparttable}
\end{table}

\subsection{Ablation Study}
\begin{figure}[htbp]
    \centering  
    \begin{minipage}{0.32\linewidth}
    \centering
    \includegraphics[width=1\textwidth]{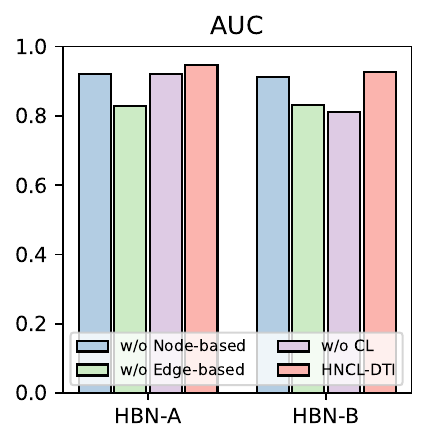}
    \end{minipage}
    \begin{minipage}{0.32\linewidth}
    \centering
    \includegraphics[width=1\textwidth]{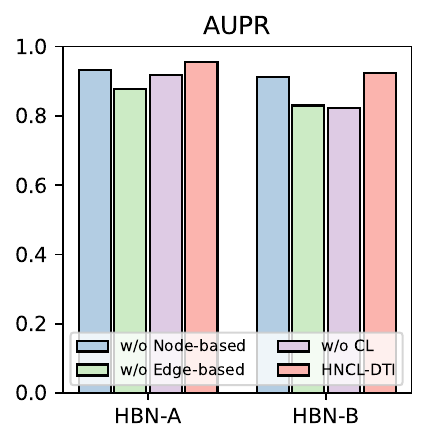}
    \end{minipage}
    \begin{minipage}{0.32\linewidth}
    \centering
    \includegraphics[width=1\textwidth]{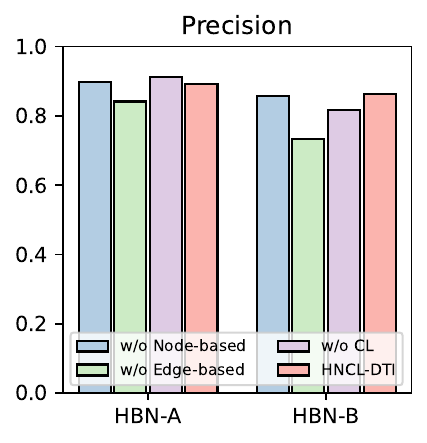}
    \end{minipage}
    %\qquad

    \begin{minipage}{0.32\linewidth}
    \centering
    \includegraphics[width=1\textwidth]{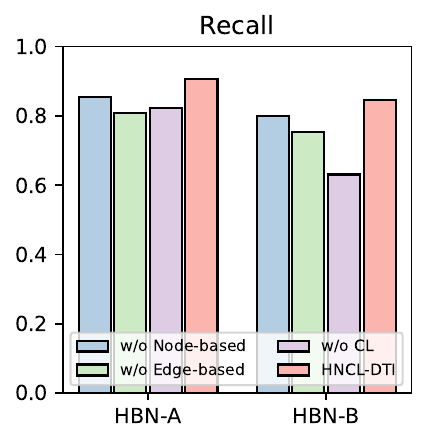}
    \end{minipage}
    \begin{minipage}{0.32\linewidth}
    \centering
    \includegraphics[width=1\textwidth]{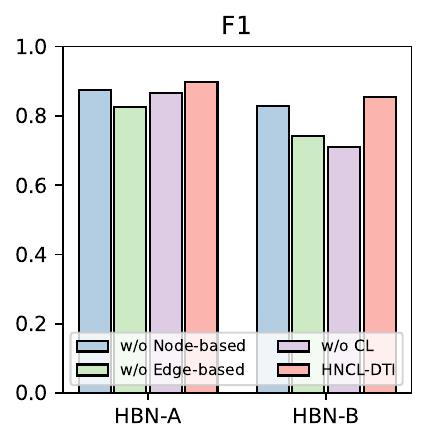}
    \end{minipage}
    \begin{minipage}{0.32\linewidth}
    \centering
    \includegraphics[width=1\textwidth]{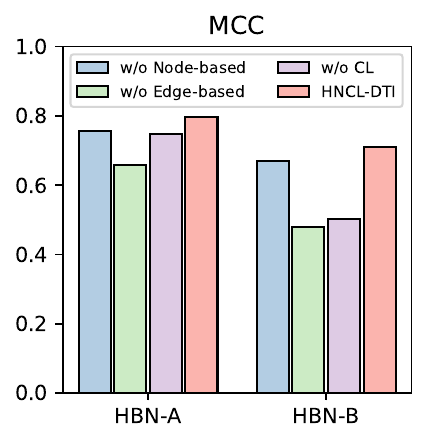}
    \end{minipage}
    \caption{\centering The ablation study of HNCL-DTI on HBN-A and HBN-B datasets.}
    \label{AS}
\end{figure}

We further perform the ablation study on different \emph{HNCL-DTI} variations to evaluate the effectiveness of each model component.
The results of the ablation study for DTI prediction on two datasets are presented in Fig.~\ref{AS}. There are three variants distinguishingly constructed:
\begin{itemize}
    \item \textbf{w/o Node-based} - This variant removes node-based attention mechanism module, that is, only edge-based attention mechanism is kept.
    \item \textbf{w/o Edge-based} - This variant removes edge-based attention mechanism module and keeps node-based attention mechanism module.
    \item \textbf{w/o CL} - This variant removes the contrastive learning module, and directly performs mean pooling for $\textbf{\emph{H}}^{node}$ and $\textbf{\emph{H}}^{edge}$ to obtain final node representations.
\end{itemize}
From the results, we can make the following three observations: (1) \textbf{w/o Edge-based} performs significantly worse than \textbf{w/o Node-based} variant. In fact, the edge-based attention mechanism incorporates edge features and node features that are not specially processed into the calculation. However, the node-based attention mechanism ignores edge features and only emphasizes node features. This proves the importance of edge relationships in the heterogeneous biomedical network.
(2) \textbf{w/o CL} performs worse than the full model. This demonstrates that contrastive learning can effectively refine both node-based and edge-based attention aggregation, leading to more effective node representations.
(3) Although all variants perform worse than the full model on the test dataset, this still verifies the effectiveness of each component of \emph{HNCL-DTI} in the DTI prediction task.

\subsection{Case Study} To evaluate the practicality of our model \emph{HNCL-DTI}, we conduct the following two studies: (i) For the HBN-A dataset, we identified 6 real drug-target pairs from~\cite{zhao2023mgdti}, which have no interaction to our benchmark test. These pairs were subsequently input into our trained \emph{HNCL-DTI} model. The model successfully predicted these 6 real drug-target pairs and these predictions were consistent with existing clinical studies and related literature. Table~\ref{CS1} gives the detailed prediction results. (ii) For the HBN-B dataset, we randomly select two drugs (\emph{i.e.}, DB00228 \& DB00408) and show their predicted targets in Table~\ref{CS2}.

\begin{table}[ht]
\caption{The case study of HNCL-DTI on dataset HBN-A.}
\vspace{-2ex}
\label{CS1}
\centering
\begin{tabular}{@{}cccc@{}}
\toprule
DrugBank ID & UniProt ID & Prediction Score & Evidence \\ \midrule
DB01050 & P37231 & 0.9999 & \cite{rinott2020ibuprofen} \\
DB01234 & O75469 & 0.9927 & \cite{tomazini2020effect} \\
DB01050 & P10415 & 0.9995 & \cite{rinott2020ibuprofen} \\
DB00608 & Q9BYF1 & 0.9863 & \cite{meo2020efficacy} \\
DB00959 & P04083 & 0.9965 & \cite{papamanoli2021high} \\
DB01050 & Q07869 & 0.9792 & \cite{rinott2020ibuprofen} \\
% DB00608 & P09488 & 0.9836 & \cite{meo2020efficacy} \\ 
\bottomrule
\end{tabular}
\end{table}

\begin{table}[ht]
\caption{The case study of HNCL-DTI on dataset HBN-B.}
\vspace{-2ex}
\label{CS2}
\centering
\begin{tabular}{@{}cccccc@{}}
\toprule
DrugBank ID & UniProt ID & Result & DrugBank ID & UniProt ID & Result \\ \midrule
DB00228 & Q06432 & True & DB00408 & P35348 & True\\
DB00228 & Q02641 & True & DB00408 & Q01959 & True\\
DB00228 & Q8WXS5 & True & DB00408 & P08172 & True\\
DB00228 & Q8N1C3 & True & DB00408 & P28221 & True\\
DB00228 & P31644 & True & DB00408 & P08912 & True\\ 
\hline
\hline
Accuracy & & 100\% & Accuracy & & 100\% \\
\bottomrule
\end{tabular}
\end{table}

\section{Conclusion}\label{Con}
This paper proposes a novel heterogeneous network-based method called \emph{HNCL-DTI} for the DTI prediction task. \emph{HNCL-DTI} utilizes two different attention mechanisms (\emph{i.e.}, node-based and edge-based) to update the feature representation of nodes by aggregating information from their connected neighborhoods, and adopts contrastive learning to adaptively learn the importance of both node-based and edge-based attention perspectives for heterogeneous biomedical graph embedding, thus improving the consistency of embedding representations. 
Experimental results on two benchmark datasets indicate that \emph{HNCL-DTI} demonstrates competitive results, and in most cases outperforms its competitors in the DTI prediction task.

\section{Acknowledgments}
This work was supported by National Key Research and Development Program of China under Grant 2023YFF1000100, the National Natural Science Foundation of China under Grant 62062063, and the Fundamental Research Funds for the Chinese Central Universities under Grant 2662023XXPY004.

\bibliographystyle{IEEEtran}
\bibliography{reference}
\end{document}